\documentclass[3p,times]{elsarticle}
\usepackage{ecrcpub}

\volume{824 (2022)}

\firstpage{136812}

\journalname{Physics Letters B}

\runauth{S.J. Brodsky, V.E. Lyubovitskij, I. Schmidt}

\jid{plb}

\jnltitlelogo{Physics Letters B}

\usepackage{amssymb,amsmath,amsfonts}
\usepackage{epsfig,graphicx,tabularx}
\usepackage{color}
\usepackage[figuresright]{rotating}

\newcommand{\seq}{\begin{subequations}}
\newcommand{\sen}{\end{subequations}}
\newcommand{\eq}{\begin{eqnarray}}
\newcommand{\en}{\end{eqnarray}}

\begin{document}

\begin{frontmatter}

\dochead{}

\title{Novel Corrections to the Momentum Sum Rule \\ for Nuclear Structure Functions} 

\author[label1]{Stanley J. Brodsky}
\ead{sjbth@slac.stanford.edu}
\author[label2,label3,label4]{Valery E. Lyubovitskij\corref{cor}}
\ead{valeri.lyubovitskij@uni-tuebingen.de}
\author[label3]{Ivan Schmidt}
\ead{ivan.schmidt@usm.cl}
\address[label1]{SLAC National Accelerator Laboratory,
Stanford University, Stanford, CA 94309, USA}
\address[label2]{Institut f\"ur Theoretische Physik, Universit\"at T\"ubingen,
Kepler Center for Astro and Particle Physics, \\
Auf der Morgenstelle 14, D-72076 T\"ubingen, Germany}
\address[label3]{Departamento de F\'\i sica y Centro Cient\'\i fico
Tecnol\'ogico de Valpara\'\i so-CCTVal, \\ Universidad T\'ecnica
Federico Santa Mar\'\i a, Casilla 110-V, Valpara\'\i so, Chile}
\address[label4]{Millennium Institute for Subatomic Physics
at the High-Energy Frontier (SAPHIR) of ANID, \\
Fern\'andez Concha 700, Santiago, Chile}
\cortext[cor]{Corresponding author at: Institut f\"ur Theoretische Physik, 
Universit\"at T\"ubingen, Kepler Center for Astro and Particle Physics, \\
Auf der Morgenstelle 14, D-72076 T\"ubingen, Germany}

\begin{abstract}

We address novel features of deep inelastic lepton scattering on nuclei at 
small Bjorken variable $x_{Bj}$. In this regime the lepton-nuclear cross section 
involves the interference between the standard lepton-quark scattering amplitude 
for the deep inelastic scattering (DIS) process on a single nucleon and 
a two-step process where diffractive scattering on a first nucleon combines 
with the amplitude for DIS on a second nucleon. 
The phases associated with the $t$-channel exchanges to the diffractive amplitude 
can produce either a destructive or constructive quantum-mechanical interference 
of the one-step and two-step amplitudes. This provides a mechanism regulating 
the respective amounts of shadowing suppression and anti-shadowing enhancement 
at low $x_{Bj}$.
Furthermore, the standard leading-twist operator product and handbag diagram analyses  
of the forward virtual Compton amplitude on the nucleus are inapplicable, 
barring a conventional probabilistic interpretation. A main observable consequence 
is the impossibility of extracting momentum and spin sum rules from nuclear structure 
functions. We present numerical predictions supporting this picture and test them against 
DIS neutrino-nucleus and charged-lepton-nucleus scattering data.

\end{abstract}


\end{frontmatter}

\clearpage 

Deep inelastic scattering (DIS) processes measure the proton structure functions, 
which can be interpreted at leading twist in QCD in terms of the probability, 
$q(x,Q^2)$, for finding a quark with a given light-front momentum fraction 
$x=\frac{k^+}{p^+}$ of the proton's light-front momentum $p^+ = p^0+p^z$, 
at large spacelike four-momentum transfer, $Q^2 = -q^2$. Longitudinal fraction 
variable $x$ is identified with Bjorken variable $x_{Bj}$ and is 
expressed in terms of invariants as $x=x_{Bj} = Q^2/(2 pq)$ with $p$ 
being the proton four-momentum.

Perturbative QCD radiative processes generate the $Q^2$ evolution of $q(x,Q^2)$, 
consistent with the renormalization group. The Mellin moments of the structure functions 
are in one-to-one correspondence with the matrix elements of the operators in the 
operator product expansion (OPE) for the forward virtual Compton amplitude 
$\gamma^* p \to \gamma^* p$. 
The second Mellin moment, in particular, gives the momentum sum rule (MSR), 
\eq
M_2 = \int_0^1 dx x \, \biggl[ 
\sum_q \, 
[q(x,Q^2) + \bar{q}(x,Q^2)] + g(x,Q^2) \biggr] = 1 \label{mom2} \,, 
\en
where $\sum_q (q+\bar{q})$ is the flavor-singlet contribution and 
$g$ is the gluon distribution. Notice that the anomalous dimensions for $M_2$ 
are equal to 0 so that the $Q^2$ dependence of the quark and gluon coefficient functions 
is guaranteed to cancel in the sum of the two contributions. 
One should stress that the gluon momentum fraction in nucleons has
to be corrected because of the diffractive contribution to DIS. 

The parton distributions in Eq.~(\ref{mom2}) are derived by Fourier transforming 
the quark-quark correlation function, written as a function of the light-cone variable, 
$z^-$, with the quark/gluon fields evaluated at light front coordinates, $z^-_{in}=0$, 
and $z^-_{out}=z^-$~\cite{Miller:2019ysh}.  
In the standard QCD scenario we can take the limit $z^- \rightarrow 0$, obtaining the integral 
definitions of the moments in Eq.~(\ref{mom2}), where the product of the two electroweak currents, 
$j^\mu(z)$ and $j^\nu(0)$ (where $j^\mu(z) = \bar\psi(z) \Gamma^\mu \psi(z)$ and 
$\Gamma^\mu = \gamma^\mu$ or $\gamma^\mu\gamma^5$), acting on 
an uninterrupted quark propagator is replaced by a local operator. 
This defines the factorized handbag diagram where the real phase of the resulting DIS amplitude -- 
the virtual Compton scattering amplitude $\gamma^* p \to \gamma^* p$ in the forward limit --  
reflects the real phase of the stable target hadron's wavefunction.
Similar results are obtained for $\bar{q}(x)$ and $g(x)$.

\begin{figure}[htb]
\begin{center}
\vspace*{-.275cm} 
\includegraphics[scale=0.25]{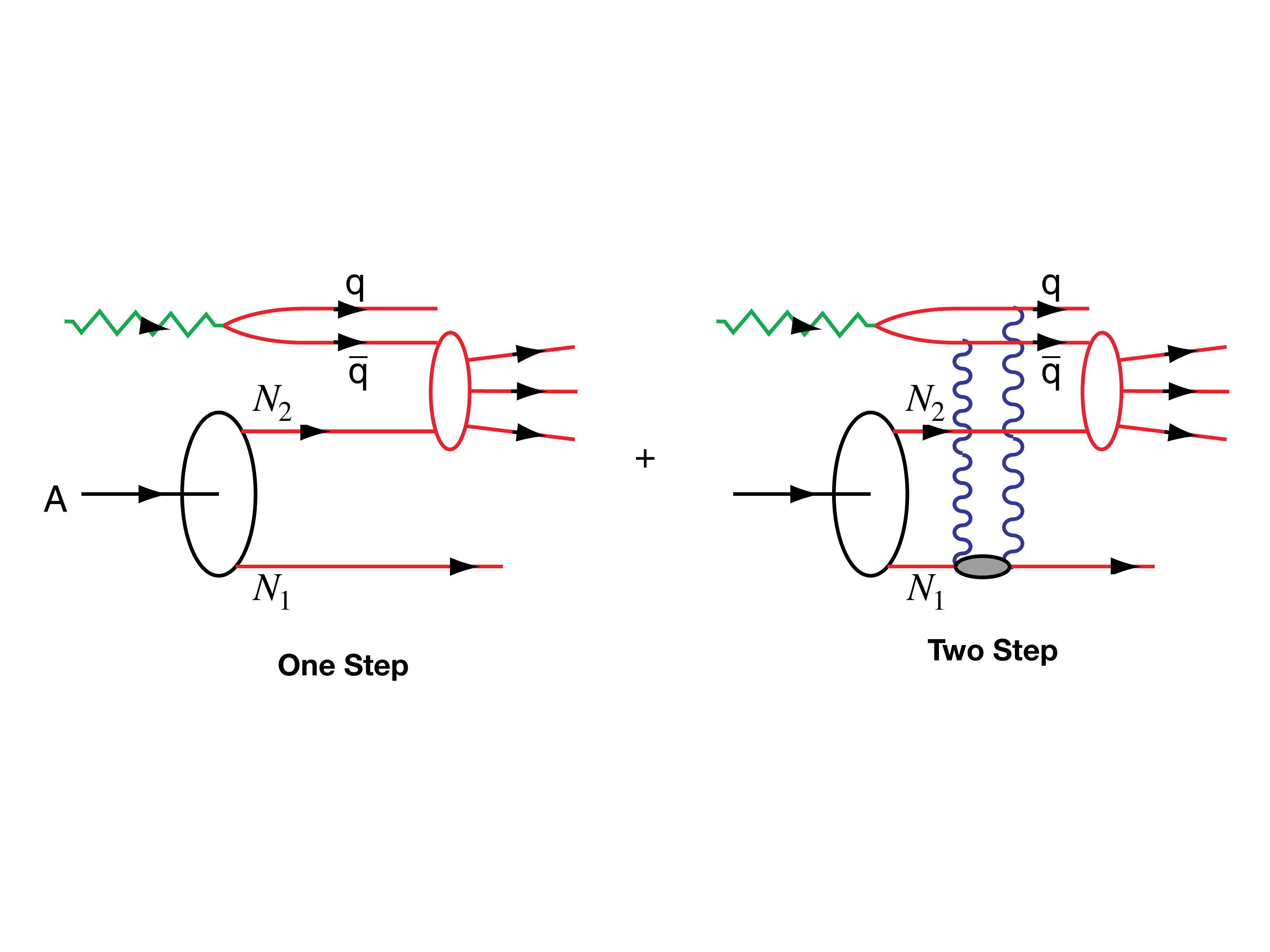}
\vspace*{-2.25cm} 
\caption{
One-step (left panel)  
and two-step (right panel) scattering amplitudes in DIS on a nucleus $A$.} 
\label{OneandTwoStep}
\end{center}
\end{figure}

In this Letter we argue that the QCD standard picture breaks down in a nucleus because 
the forward virtual Compton scattering amplitude for the process,  
$\gamma^*(Q^2) A \to \gamma^*(Q^2) A$, includes at low $x$ a leading-twist contribution from 
the interference between the one-step and two-step amplitudes shown in Fig.~\ref{OneandTwoStep}. 
In particular, the initial scattering in the two-step amplitude on the front-face nucleon $N_1$ is 
Diffractive DIS (DDIS): $\gamma^* N_1 \to [q \bar q] N_1^\prime$ which leaves $N_1$ intact. 
The propagating vector $(q \bar{q})$ system then interacts inelastically on $N_2$:  
$[q \bar q]+ N_2 \to X$. 
The two step amplitude interferes with the one-step amplitude $\gamma^*+ N_2 \to X$ on $N_2$.  
The interior nucleon $N_2$ sees two fluxes, the virtual photon $\gamma^*$ and the secondary 
beam $(q\bar{q })$ generated by DDIS on $N_1$. In effect, nucleon $N_1 $ ``shadows" $N_2$. 
The smooth limit, $z^-\rightarrow 0$,  is inapplicable in this case since the currents can act 
on different nucleons.   
Because of this type of quantum interference between the scattering amplitudes from two-step 
and one-step amplitudes in nuclei, incoherent scattering on a single quark is ruled out by definition, 
as well as its probabilistic interpretation and the validity of the MSR, Eq.~\eqref{mom2}. 
The contribution to doubly virtual Compton scattering on a nucleus $\gamma^* A \to \gamma^* A$ 
from the interference of two-step and one-step amplitudes cannot be reduced to a handbag amplitude 
where two currents interact on an interrupted quark propagator.

The picture presented here is consistent with the analysis of shadowing in the dipole formalism.  
The dipole analysis uses the target rest frame where two length scales are important.  
One is the coherence length ($\approx 1/\sqrt{Q^2}$), which has to be much larger than  
the inter-nucleon separation, and the other is the $\bar{q} q $ transverse separation,   
which has to be large enough so that the virtual photon interacts with a sizable cross section.  
These large length scales conflict with the applicability of the OPE in nuclear shadowing processes, 
and therefore, on the validity of the nuclear MSR~\cite{Brodsky:2019jla}.   

\begin{figure}
\begin{center}
\vspace*{-1.75cm}
\includegraphics[scale=0.4]{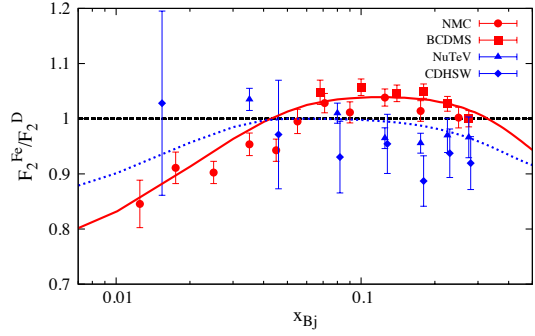}
\vspace*{-5cm}
\caption{Comparison of the ratio of iron to deuteron nuclear structure functions 
measured in the DIS neutrino-nucleus scattering (NuTeV~\cite{Tzanov:2005kr}, 
CDHSW~\cite{Berge:1989hr}), and muon-nucleus scattering (BCDMS~\cite{Benvenuti:1987az} 
and NMC~\cite{Amaudruz:1995tq,Arneodo:1996rv}). 
Blue dotted and red lines correspond to our numerical predictions for
the DIS neutrino-nucleus and charged-lepton-nucleus scattering data. 
All data are displayed in the online Durham HepData Project
Database~\cite{HepData}. 
Anti-shadowing is absent in the neutrino charged current data.}
\label{Comparison}
\end{center}
\end{figure}

The breaking of the MSR, according to the quantum interference interpretation 
that we suggest here, and the flavor dependence of the $t$-channel exchanges in the diffractive 
process, also provide a much sought-for explanation of one of the most surprising results of the 
NuTeV measurement~\cite{Tzanov:2005kr} of nuclear structure functions in the DIS 
charged-current reactions $\nu A \to \mu X$, concerning the absence of anti-shadowing 
in the domain $0.1 < x_{Bj}< 0.2$ (see Fig.~\ref{Comparison}). NuTeV's initial measurement   
was further substantiated by the global analysis conducted by nCTEQ~\cite{Schienbein:2007fs} 
where, based on the fully correlated covariant error matrix provided in~\cite{Tzanov:2005kr}, 
it is shown that the tension between the  $\nu Fe$ and  $\mu Fe$ data 
does not allow for a ``compromise'' fit that includes both sets.

A similar conclusion has also been reached in the 
more recent experimental analyses of Refs.~\cite{Mousseau:2016snl,Tice:2014pgu,Kovarik:2011dr}, 
as well as in Ref.~\cite{Kalantarians:2017mkj}, where following an accurate analysis of the 
$Q^2$ dependence of the different data sets, the nuclear parton distribution functions (PDFs) 
measured in the DIS neutrino reactions 
are shown to have no anti-shadowing enhancement and are thus distinctly different from 
the corresponding PDFs measured in the charged lepton DIS (see Fig.~\ref{Comparison}). 
More quantitative analyses are in currently in progress to better constrain the shadowing and 
anti-shadowing regions.

The striking difference between neutrino vs. charged lepton DIS measurements is in direct conflict 
with the conventional expectation that the quark and gluon distributions of the nucleus are 
universal properties of the nuclear eigenstate and are thus process independent (see discussion 
in~\cite{Kopeliovich:2012kw}). Moreover, the NuTeV measurement contradicts
the expectation that anti-shadowing $\frac{d\sigma_A}{A d\sigma_N} < 1$ which is observed 
in the domain $0.1 < x_{Bj}< 0.2$ in order to restore the MSR~\cite{Nikolaev:1975vy}. 

Understanding the workings of low $x$ PDFs in nuclei and their impact on the MSR 
is of the utmost importance, especially as more precise measurements on nuclear PDFs 
will be made available at the upcoming Electron Ion Collider (EIC).

We describe a scenario by which both shadowing and anti-shadowing originate as Glauber phenomena 
involving the constructive vs. destructive interference of two-step and one-step amplitudes 
illustrated in Fig.~\ref{OneandTwoStep} of Ref.~\cite{Stodolsky:1994ka}.
The first step of the two-step amplitude involves leading-twist DDIS 
on a front-face nucleon $N_1$, which leaves the nucleon intact.  
DDIS underly the shadowing and antishadowing of nucleus structure functions through 
the two-step amplitude. DDIS in $\gamma^* N \to N X$ 
reactions has been observed to satisfy Bjorken scaling, and approximately 10\% of high energy DIS events 
are diffractive~\cite{Derrick:1993xh,Ahmed:1994nw}. In the Regge theory of strong interactions diffraction 
occurs  through the exchange of either a Pomeron or a Reggeon trajectory. In quantum chromodynamics (QCD), 
the Pomeron and the Reggeon correspond to two gluons and to quark-antiquark color-singlet exchanges, 
respectively. The diffractive process is leading twist and it, therefore, displays Bjorken scaling.  

The second step of the two-step amplitude is a standard inelastic scattering on a second nucleon, $N_2$, 
producing a final state $X$. The interference of the two-step amplitude with the DIS event on nucleon $N_2$ 
shown in Fig.~\ref{OneandTwoStep} (bottom panel) can produce shadowing or anti-shadowing of the nuclear 
PDF depending on the phase of the DDIS amplitude.  

Unlike the handbag diagram, the phase of the deeply virtual amplitude arising from the Glauber interference 
amplitudes is always complex. In a nucleus we define the ratio of structure functions, $R_A=F_2^A/A F_2^N$, 
in terms of the imaginary part of the forward quark-nucleon scattering amplitude, $T_{qN}$ as,
\eq 
R_A(x,Q^2) = \frac{\int ds \, dk_T^2 \, 
\Im m T_{qA}(s,k^2)}{A \, \int ds \, dk_T^2 \, \Im m T_{qN}(s,k^2)} 
\en
with $T_{qA}$ given in Glauber theory as follows~\cite{Brodsky:1989qz,Brodsky:2004qa},
\eq 
T_{qA} &=&  \sum_{j=1}^A \frac{1}{j} \binom{A}{j} T_{qN} 
\left(\frac{i T_{qN}}{4 \pi p_{CM} s^{1/2} (R^2 + 2b)} \right)^{j-1} \nonumber\\ 
 &\approx & A T_{qN}\left(1 +  \frac{i (A-1) T_{qN}}{8 \pi p_{CM} s^{1/2} (R^2 + 2b)} \right)
 \label{eq:TqA}
\en 
where $s=(k+p)^2\approx 1/x$ is the parton-proton center of mass energy squared, 
$k^2$ is the quark virtuality, $k_T^2$ is the quark transverse momentum squared, 
related linearly to $k^2$, $p_{CM}$ is the quark-proton center of mass momentum, 
$R=1.12 \, A^{1/3}$ fm is the nuclear radius, $b = 10$ GeV$^{-2}$ is the parameter 
defining the slope of the non forward amplitude, $T_{qN}= T_{qN}(s,k^2)\exp({-b q_T^2})$. 
$T_{qN}$ describes Regge exchanges with all allowed $J^{PC}$ quantum numbers. 
We used a form containing the two essential contributions from Pomeron ($\beta_1$) 
and Reggeon ($\beta_{1/2}$) exchanges, respectively given by~\cite{Brodsky:1989qz,Brodsky:2004qa}, 
\eq 
\hspace*{-0.75cm} 
T_{qN} \approx \sigma \left(i s \beta_1(-k^2) 
+ \frac{1}{\sqrt{2}} (1-i) s^{1/2} \beta_{1/2}(-k^2) \right)
\label{eq:TqN} \,. 
\en
From this expression one can clearly see that multiplying by the phase $i$ from the propagating 
intermediate state, or Glauber cut, the relative phase of the two-step amplitude, 
$\propto i \times i$ is destructive if the diffractive component is due to Pomeron exchange 
thus producing shadowing. On the contrary, Reggeon exchanges enable constructive interference, 
$\propto i \times (-i)$, thus anti-shadowing. 
Notice that due to the inverse proportionality of $s$ and $x$, the anti-shadowing term is predicted 
to appear at larger values of $x$. The resulting effect from the constructive interference appears in the 
$0.1 < x_{Bj} < 0.2$, domain of the nuclear PDF. 
Notice that the exchange of the same Reggeon also leads to the Kuti-Weisskopf prediction: 
$F_{2}^p(x,Q^2) -  F_{2}^n(x,Q^2) \propto \sqrt{x}$ (this result is consistent with recent evaluations 
in Refs.~\cite{Harland-Lang:2014zoa,Alekhin:2015cza}). 

Eqs.~(\ref{eq:TqA}) and~(\ref{eq:TqN}) describe the situation in Fig.~\ref{OneandTwoStep} 
where $N_1$ is the front-face nucleon and $N_2$ is an interior nucleon. 
In the one-step process only $N_2$ interacts via Pomeron exchange, while $N_1$ does not. 
One can see that if the scattering on $N_1$ is, {\it e.g.}, via Pomeron exchange, and both 
amplitudes have different phases, diminishing the $\bar{q}$ flux that reaches $N_2$. 
The interior nucleon, $N_2$, thus sees two fluxes -- the incident virtual photon $\gamma^*$ 
and the $q \bar{q}$ vector system, $V^0$ (grey blob in the bottom rungs in Fig.~2), produced 
from DDIS on $N_1$. The relative phase of the one-step and two-step amplitudes is the critical 
factor of $i$ from the Glauber cut times the phase of Pomeron exchange in DDIS.  
The destructive interference is why $N_2 $ does not see the full flux -- it is shadowed by $N_1$. 
Thus shadowing of the nuclear PDF is due to additional physical, causal events within the nucleus.  
Being defined by interference terms, the MSR can fail for the nuclear PDFs: 
as shown in our numerical evaluation in Fig.~\ref{Comparison}, shadowing and anti-shadowing 
do not need to compensate each other to restore the MSR. 
Our analysis is based on general principles. We note that our approach does work well describing 
measurements of deep inelastic lepton scattering on nuclei at small Bjorken variable $x_{Bj}$.
The parameters in the formulae~(\ref{eq:TqA}) and~(\ref{eq:TqN}) have a typical uncertainty of say 10\%.

Thus unlike shadowing, anti-shadowing from Reggeon exchange is flavor specific; {\it i.e.}, 
each quark and anti-quark will have distinctly different constructive interference patterns.  
The flavor dependence of anti-shadowing explains why its amount varies in electron 
(neutral electromagnetic current) vs. neutrino (charged weak current) DIS reactions, 
based on the flavor composition of the DIS isoscalar nuclear and nucleon structure functions, 
\begin{subequations}
\eq
\frac{1}{2x} F_2^{\nu N(A)} &=& 
d_{N(A)} + s_{N(A)} + \bar{u}_{N(A)} + \bar{c}_{N(A)} + \ldots \,,\\
\frac{1}{x} F_2^{\ell N(A)} &=& 
  \frac{4}{9} \Big(u_{N(A)} + \bar{u}_{N(A)}\Big) 
+ \frac{1}{9} \Big(d_{N(A)} + \bar{d}_{N(A)}\Big) 
+ \frac{1}{9} \Big(s_{N(A)} + \bar{s}_{N(A)}\Big) + \ldots \,, 
\en
\end{subequations}
where $\sigma_A/A\sigma_N \approx {F_2^A}/{A F_2^N}$.

Notice that $V^0$ propagates on-shell. 
This means that not all the propagators in the graph can be considered as being hard (of order $Q^2$): 
this invalidates the OPE, and, as a consequence, the MSR. 
Most important, the finite path length due to the on-shell propagation of $V^0$  
between $N_1$ and $ N_2$ contributes to the distance $({\Delta z})^2$ between the two virtual photons 
in the $\gamma^*A \rightarrow \gamma^* A$ amplitude. One no longer has $({\Delta z})^2 \approx {1/Q^2}$.  
The distance between the currents cannot be less than the inter-nucleon distance, invalidating also 
the OPE and the parton MSR. 

The one-step two-step mechanism describing nuclear shadowing and anti-shadowing 
explains how parton reinteractions in a nucleus are an essential element in high energy reactions, 
since they can generate non-trivial contributions at leading twist and thus survive at high $Q^2$ 
and high invariant mass, $W^2 = (q+p)^2.$ Most importantly, it brings to the forefront the importance 
of describing deeply virtual scattering phenomena from nuclei at the amplitude level, where the phase 
structure plays a key role.

\begin{figure}
\begin{center}
\vspace*{-.4cm}
\includegraphics[scale=0.2]{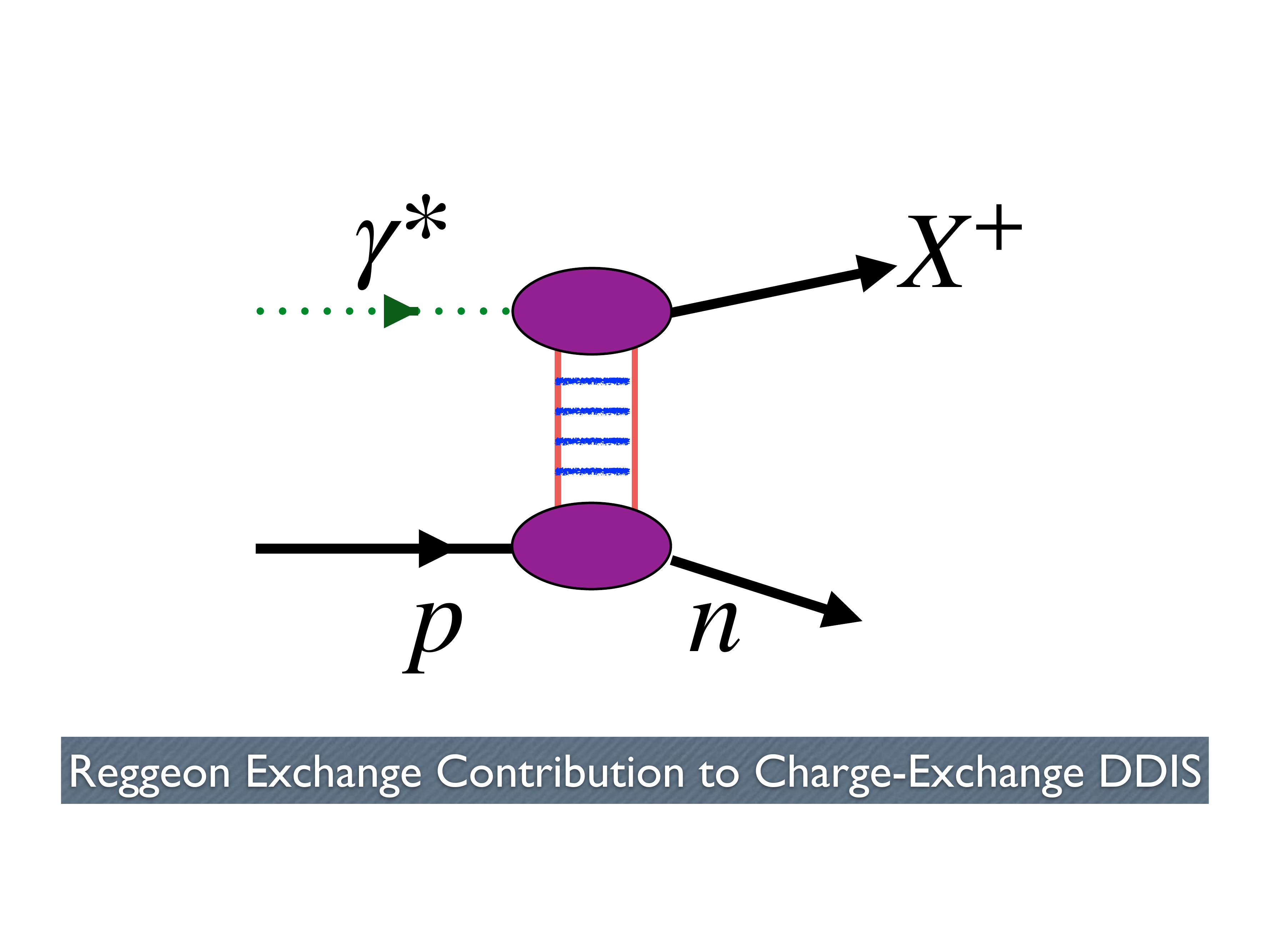}
\vspace*{-1cm}
\caption{QCD mechanism for charge-exchange leading-twist DDIS $\gamma^* p \to n X^+$.}
\label{CEDDIS}
\end{center}
\end{figure}

In order to test the explanation of anti-shadowing, one could verify the existence of Bjorken-scaling,  
leading-twist charge exchange DDIS reactions such as  $\gamma^* p \to n X^+$, with a rapidity gap due 
to an $I=1$ Reggeon exchange. Here $X^+$ is the sum of final states with charge $Q= 1$.
Since Pomeron exchange does not contribute to the charge exchange process, this would single out  
Reggeon exchanges as the source of anti-shadowing. This process is shown in Fig.~\ref{CEDDIS}.
Furthermore, deeply virtual Compton scattering (DVCS) and related experiments on nuclei would display 
a similar shadowing/anti-shadowing pattern as in the forward DIS limit, enhanced by the extra 
four-momentum, $t$, dependence~\cite{Liuti:2005gi}.

Note that other classes of DIS processes have been shown to be directly sensitive to the rescattering 
of the struck quark, for example, the pseudo-$T$-odd Sivers effect~\cite{Brodsky:2002cx,Brodsky:2013oya}.  
The ``handbag" approximation to DVCS defines the ``static"  
contribution~\cite{Brodsky:2008xe,Brodsky:2009dv} to the measured PDFs, Transverse Momentum 
Distributions (TMDs), etc. 
Similarly, nuclear DDIS involves the exchange of gluons after the quark has been struck by the 
lepton~\cite{Brodsky:2002ue}. In each case the corresponding scattering amplitude is not given by the 
handbag diagram, since interactions between the electroweak currents at different points, $j^\mu(z)$ 
and $j^\nu(0)$, are essential. The FSI associated with the Sivers effect (``lensing" 
corrections~\cite{Brodsky:2002cx,Burkardt:2003uw}) survive when both $W^2$ and $Q^2$ are large, 
since the vector gluon couplings grow with energy.  
However, in this case the final state phase is associated with a Wilson line which, 
at variance with the nuclear two-step process, produces an augmented Light Front Wave 
Function (LFWF)~\cite{Brodsky:2010vs}, that does not affect the $x$ moments, or the sum rules.  

Even in the case of the proton, Mueller~\cite{Mueller:1996hm} has noted that the OPE applied to DIS 
fails at small $x_{Bj}$. The mechanism is however different: due to the diffusion of gluons 
to small values of momentum transfer, it is not possible to separate soft and hard scales 
in this region. This might be related to the fact that at small $x$ there are Fock states with 
a large number of small $x$ partons, due to processes such as parton fusion and overlapping, 
which means that the invariant mass of these configurations is undefined. Therefore the usual 
derivation of sum rules based on the handbag diagram for (forward) double virtual Compton 
scattering may be inapplicable even on a single nucleon (see detailed discussion in 
Ref.~\cite{Brodsky:2021jmj}). 
In particular, DDIS does not satisfy the OPE nor the MSR. 
The failure of the OPE and the MSR for DDIS implies that shadowing and antishadowing 
will not compensate each other to restore the MSR for nuclear structure function. 
The NuTeV results provide direct evidence for the failure of the MSR for the nuclear 
structure functions. 

In conclusion, we summarize the main results of our paper: 

(1) We have illustrated why anti-shadowing of nuclear structure functions is 
non-universal, {\it i.e.}, flavor dependent, and why shadowing and anti-shadowing phenomena 
are incompatible with the standard application of the OPE.  
As a consequence, sum rules cannot be extracted from nuclear DIS structure functions.  

(2) We reiterate that because of the rescattering dynamics, the DDIS amplitude acquires a complex 
phase from Pomeron and Regge exchange;  thus final-state  rescattering corrections lead to  
nontrivial ``dynamical" contributions to the measured PDFs, {\it i.e.}, they are a consequence 
of the scattering process itself~\cite{Stodolsky:1994ka,Brodsky:1989qz,Brodsky:2013oya,Bauer:1977iq}. 
The $I = 1$ Reggeon contribution to DDIS on the front-face nucleon then leads to flavor dependent 
anti-shadowing~\cite{Brodsky:1989qz,Brodsky:2004qa}. This could explain why the NuTeV charged 
current measurement $\mu A \to \nu X$ scattering does not appear to show anti-shadowing, 
in contrast to deep inelastic electron-nucleus scattering as discussed in 
Ref.~\cite{Schienbein:2007fs} and illustrated in Fig.~\ref{Comparison}. 

(3) Finally, the Regge exchanges-based theoretical description of shadowing-antishadowing 
presented here is ideal for realistic, quantitative evaluations of the nuclear structure 
functions in the upcoming EIC kinematic framework.  

\section*{Acknowledgments}

We are thankful to Simonetta Liuti for her collaboration 
in earlier stages of this work. 
This research was supported by the Department of Energy  
contract DE--AC02--76SF00515 (SJB), by ANID PIA/APOYO AFB180002 (Chile) 
by FONDECYT (Chile) under Grants No. 1180232 and No. 1191103, and 
by Millennium Institute for Subatomic Physics
at the High-Energy Frontier (SAPHIR) of ANID, Code: ICN2019\_044 (Chile).  
SLAC-PUB-17626.

\end{document}